\documentclass[11pt]{article}
\usepackage{hyperref}
\hypersetup{
pdftitle={Report on a Boston University Conference December 7-8, 2012 on `How Can the History and Philosophy of Science Contribute to Contemporary U.S. Science Teaching?'},
pdfauthor={Peter Garik, Yann Ben\'etreau-Dupin},
colorlinks=true,
linkcolor={black},
citecolor={blue},
urlcolor={blue},
}
\usepackage{setspace,commath,amsfonts,natbib,geometry,lastpage,fancyhdr,wrapfig,graphicx}
\let\quoteOLD\quote
\def\quote{\quoteOLD\small\singlespacing}
\author{Peter Garik\footnote{School of Education, Two Silber Way, Boston University, Boston, MA 02215, USA, \href{garik@bu.edu}{garik@bu.edu}} \& Yann Ben\'etreau-Dupin\footnote{Department of Philosophy and Rotman Institute of Philosophy, Western University, London, Canada, N6A 5B8, \href{ybenetre@uwo.ca}{ybenetre@uwo.ca}}}
\title{Report on a Boston University Conference December 7-8, 2012 on
`How Can the History and Philosophy of Science Contribute to Contemporary U.S. Science Teaching?'\footnote{Published in \emph{Science \& Education}, September 2014, 23(9): 1853--1873. \href{http://dx.doi.org/10.1007/s11191-014-9716-8}{doi:10.1007/s11191-014-9716-8}.}}
\date{}

\begin{document} 
 
\maketitle

\begin{abstract}
This is an editorial report on the outcomes of an international conference sponsored by a grant from the National Science Foundation (NSF) (REESE-1205273) to the School of Education at Boston University and the Center for Philosophy and History of Science at Boston University for a conference titled: How Can the History and Philosophy of Science Contribute to Contemporary U.S. Science Teaching? The presentations of the conference speakers and the reports of the working groups are reviewed. Multiple themes emerged for K-16 education from the perspective of the history and philosophy of science. Key ones were that: students need to understand that central to science is argumentation, criticism, and analysis; students should be educated to appreciate science as part of our culture; students should be educated to be science literate; what is meant by the nature of science as discussed in much of the science education literature must be broadened to accommodate a science literacy that includes preparation for socioscientific issues; teaching for science literacy requires the development of new assessment tools; and, it is difficult to change what science teachers do in their classrooms. The principal conclusions drawn by the editors are that: to prepare students to be citizens in a participatory democracy, science education must be embedded in a liberal arts education; science teachers alone cannot be expected to prepare students to be scientifically literate; and, to educate students for scientific literacy will require a new curriculum that is coordinated across the humanities, history/social studies, and science classrooms.
\end{abstract}

\section{Introduction}

This report is meant to introduces the reader to the objectives and outcomes of an international conference sponsored by a grant from the National Science Foundation (NSF) (REESE-1205273) to the School of Education at Boston University and the Center for Philosophy and History of Science at Boston University. The full title of the conference as offered in the proposal for the conference funding was \emph{How Can the History and Philosophy of Science Contribute to Contemporary U.S. Science Teaching?} A proposal for an interdisciplinary conference and workshop to formulate research plans to study the theory, curriculum and pedagogy for using the history and philosophy of science in science education. The articles that appear in this issue by Douglas Allchin and Gerald Holton, and the article by David W. Rudge, David P. Cassidy, Janice M. Fulford, and Eric M. Howe, relate to this conference.

What follows is divided into six sections:
\begin{list}{-}{}
\item a description of the conference format (\S~\ref{Format});
\item the rationale and objectives for the conference as presented to the National Science
Foundation in the conference grant proposal (\S~\ref{Rationale});
\item an overview of the presentations made at the conference (\S~\ref{Summary});
\item an overview of the reports of the working groups that formed at the conference (\S~\ref{Groups});
\item a discussion of the principal outcomes of the conference as interpreted by the
conference organizers (\S~\ref{Discussion}); and,
\item conclusions drawn from the presentations, working group reports, and the analysis in the discussion (\S~\ref{Conclusions}).\\
\end{list}

\subsection{Editorial Privilege}

In writing the following, we have adopted the editorial privilege of the conference organizers to offer the reader interpretations and conclusions that we feel can be safely drawn in the spirit of the presentations. This editorial and our interpretations have \emph{not} been reviewed by the conferees.

\section{Conference Format and Description}
\label{Format}

The conference was held at Boston University on December 7th and 8th in 2012, as one of the colloquia of the 53rd Annual Boston Colloquium for the Philosophy of Science hosted by the Center for Philosophy and History of Science (CPHS) at Boston University.
The first day of the conference was a public conference hosted by the CPHS and the School of Education. It was an all-day event with three panels of speakers in the morning and three panels in the afternoon. On the second day of the conference, the conferees divided themselves among working groups to prepare reports on the status of the history and philosophy of science (HPS) in science education. The 75 invitees represented a broad spectrum of expertise ranging across science content, science education research, social studies, education research, history of education, project evaluation, and cognitive theory, with science historians, philosophers, and science teachers contributing. Among the invitees were many members of the International History, Philosophy and Science Teaching Group (\href{http://ihpst.net/}{http://ihpst.net/}). To involve the next generation of HPS educators, graduate students from Boston University and from Western University (Canada) were invited to participate and assist in keeping records of discussions.\\

\subsection{Selection of Presentations}

At the suggestion of the NSF, conference invitees represented a wide range of educational expertise: cognition, evaluation, science content, and science education. The Center for Philosophy and History of Science has a long history of sponsoring conferences on the philosophy of science. Corresponding to this sponsorship, philosophers of science were invited who possessed an interest in science education. Arrangement of topic panels was done iteratively. An initial set of potential topics was selected by the conference organizers. Based on the availability of speakers, a final set of panel titles was selected. Through this method emerged a set of panels and speakers who covered many of the initially intended topics, but also topics not originally included. Inevitably, with such a rich cast of conferees, not everyone could be represented on a panel. To accommodate a few additional topics, plenary talks were also invited.

The final panels were:
\begin{list}{-}{}
\item \emph{History of HPS in Science Education}\\
Speakers: Gerald Holton (Harvard University), Michael Matthews (University of New South Wales). Comments by Sevan Terzian (University of Florida).
\item \emph{HPS and the Science Frameworks}\\
Speakers: Richard Duschl (Pennsylvania State University and NSF), Gregory Kelly (Pennsylvania State University). Comments by Michael Ford (University of Pittsburgh), Jacob Foster (Massachusetts Department of Education), Katherine McNeill (Boston College).
\item \emph{Teaching and Learning with HPS I: Outcomes for Teachers and Students}\\
Speakers: Fouad Abd-El-Khalick (University of Illinois at Urbana-Champaign), Fanny Seroglou (Aristotle University of Thessaloniki). Comments by Katherine Brading (University of Notre Dame), Ricardo Lopes Coelho (University of Lisbon).
\item \emph{Teaching and Learning with HPS II: Outcomes for Teachers and Students}\\
Speakers: Xiaodong Lin-Sielger (Columbia University), Douglas Allchin (University of Minnesota). Comments by Frank Keil (Yale University), Michael Clough (Iowa State University).
\item \emph{Using HPS in the Classroom: Ethical Reasoning and Modeling}\\
Speakers: Mildred Solomon (The Hastings Center and Harvard Medical School), Tina Grotzer (Harvard University). Comments by Jeanne Chowning (Northwest Assocation for Biomedical Research), Luciana Garbayo (University of Texas at El Paso).
\item \emph{HPS in K-12 Professional Development}\\
Speakers: Dietmar Höttecke (Universität Hamburg), David Rudge (Western Michigan University). Comments by John Clement (University of Massachusetts Amherst), Barbara Crawford (University of Georgia).
\end{list}

Plenary presentations were made by Robert Bain (University of Michigan), David Klahr (Carnegie-Mellon University), Patrick Morris (Menagerie Theatre Group), John Stachel (Boston University). In addition, there was a plenary panel of philosophers with Alisa Bokulich (Boston University), Katherine Brading (University of Notre Dame), Carol Cleland (University of Colorado Boulder), Patrick Forber (Tufts University), Luciana Garbayo (University of Texas at El Paso), and Christopher Lehrich (Boston University).

\section{Conference Rationale and Objectives}
\label{Rationale}

The ambitious title of the conference was meant to reflect a need to address the widely held belief that there is positive value in teaching HPS in support of science education. It is a belief supported by both the American Association for the Advancement of Science in \emph{Science for All Americans} \citep{AAAS1990} and the National Academy of Science in the \emph{National Science Education Standards} \citep{NRC1996} and the more recent \emph{Framework for K-12 Science Education} (\emph{Framework}) \citep{NRC2011}. Historically, this belief in the value of teaching the history of science pre-dates all of these standards \citep{Conant1957,Holton1985,Matthews1994}. Nevertheless, the hypothesis that teaching the history and philosophy of science has practical value in the classroom and for meeting national science education standards does not have a strong evidential basis. As summed up by \citet{Teixeira2009a}, ``there is an urgent need to assess the efficiency of HPS in science teaching in the classroom, especially in relation to conceptual learning, opinions and attitudes toward the nature of science, argumentation and meta-cognition.''

It was the gulf between the assumed wisdom and the research base to support the integration of HPS into the classroom that motivated the conference. The principal outcome of the conference was expected to be a sequence of next research and development steps to be taken to close the gap between what wisdom suggests and evidence supports. To work towards this goal, the Center for Philosophy and History of Science and the School of Education brought together scholars in the history and philosophy of science, scientists, educators, and education researchers to set a research agenda for testing the value of teaching HPS as part of the science curriculum.

Specifically, the objectives and projected outcomes of the meeting were:

\emph{1. A curriculum-centered research agenda that distinguishes between different aspects of HPS and their different roles in the support of student learning of science, engineering and citizenship.}

The rationale for this objective followed from different aspects of HPS all having possible roles to play in the K-16 science curriculum. Different developers in the HPS community appear focused on different aspects of HPS. For example, \citet{Allchin2011,Allchin2012} emphasizes interesting human stories and the sociocultural impact of science with a ``whole science'' approach appropriate for general citizenship, while the work of the European History and Philosophy in Science Teaching group (\href{http://hipst.eled.auth.gr/}{http://hipst.eled.auth.gr/}) emphasizes the role of HPS for instruction in science epistemology and for the understanding of scientific concepts \citep{Coelho2013}. This latter approach may be closer to the Framework with its call for increased emphasis on modeling. From a teaching and learning perspective, it is not always clear when HPS has more relevance to student learning in the science classroom as opposed to the history or social studies classroom.
In keeping with this objective, a hoped for outcome of the conference was to be a clarification of the different roles for HPS and their places in the general K-16 curriculum.

\emph{2. A student-centered research agenda that focuses on the impact of HPS on student learning.}

In the United States and internationally, there have been many small scale studies of the
impact of including an aspect of HPS in a science classroom. A synthesis and critique of prior results in service of developing a theory of how to integrate HPS into a pedagogy to improve student science mastery (content and understanding) is needed.

Another hoped for outcome of the conference was to be the definition of a research agenda to achieve this objective, along with recommendations of which levels and topics of HPS might result in the most productive initial research to shape future efforts.

\emph{3. A teacher-centered research agenda to determine professional development objectives for preparing science teachers in HPS.}

Corresponding to research on supporting student learning with HPS, there must be an equal effort to prepare teachers to integrate HPS in the classroom. Teachers teach the way they have been taught \citep{Desimone2002}. If science teachers are to use aspects of HPS in the classroom, professional development must be offered that leads to teachers' integrating HPS into their pedagogical content knowledge \citep{Shulman1987}. In their analysis of physics education, \citet{Hottecke2011} have identified obstacles to induce teachers to adopt HPS in their classrooms. These include a culture of teaching science that is different from that of teaching other subjects; teachers' beliefs about teaching science; lack of clarity in curricular standards on the role of HPS; and, lack of HPS appropriate content in textbooks. Although theirs was an international study and focused on physics, there is little reason to doubt that the barriers are similar in the U.S., and that the barriers apply to all of the sciences. If the student-centered research leads to the conclusion that HPS enriches students' science learning, and helps achieve more general curricular objectives, the development of professional development standards and methods must be seen as necessary.

It was further hoped that an outcome of the conference would be a report on the barriers to teachers' adoption of HPS in the U.S. and necessary steps to overcome these barriers.

\section{Summary of Conference Presentations}
\label{Summary}

A summary review of the public talks follows here that highlights presenters' major themes. The titles of the talks, accompanying presentations, and video of the speakers' presentations are available on the conference website, \href{http://www.bu.edu/hps-scied/}{http://www.bu.edu/hps-scied/}. To avoid redundancy, in some instances remarks by conferees are referenced in the Discussion and Conclusion sections.

Throughout the presentations and working group discussions, there were frequent references to the ``nature of science'', often denoted as NOS in the science education literature. A group of philosophers and historians of science present at the conference characterized NOS as ``the marshaling of empirical evidence in support of a claim as a basis for action,'' and added that ``different domains of science use different strategies and methods for doing this.'' However, the definition of the nature of science and how it should be framed in terms of teaching goals is elusive in the science education literature. For the purposes of this conference review, for science education we adopt the view of \emph{Science for All Americans} \citep{AAAS1990} that the nature of science has three principal components, a scientific world view, scientific methods of inquiry, and the nature of the scientific enterprise. We refer the reader to Chapter 1 of \emph{Science for All Americans} for elaboration for a general understanding of what the target concepts are. Each of the conferees is likely to have a variant of this view of the NOS. The reader is referred to their papers for these views, and to \citep{Duschl2013}.

Another slippery term is science literacy. In this report we will interpret science literacy operationally. In a democracy, one of the principal objectives of an education (especially a public education) is to prepare students to be socially and politically contributing citizens. With this in mind, we take it that the scientifically literate citizen has sufficient science content knowledge and understands how science is conducted sufficiently well to be able to make personal decisions relating to the impact of science on their own lives, and decisions at the ballot box about science policy that affects society.

Gerald Holton opened the conference with a reminiscence of the history of Project Physics \citep{Holton1967,Holton2003,Rutherford1981}, a recollection and an exhortation that encapsulates the theses and conclusions of the conference and workshops. Professor Holton placed science education firmly within a liberal arts education: science education is part of education in human culture, and an approach to science education that links the humanities and the sciences can engage a broader cross-section of students. Holton cited Project Physics as an example of a humanistic approach to the physics curriculum which was shown through extensive research \citep{Ahlgren1973,Welch1973,Welch1972} to have improved students' attitudes towards science while successfully teaching physics content.

The view that such a humanistic approach to science contributes to students' engagement and to both an understanding of science content and an understanding of science as part of modern culture led Holton to say that ``an educator of young persons has the ethical imperative to try to convey a way of thinking and functioning, a view of the world in which sciences has its proper place. It must be part of some preparation for a long life in a society in which science and technology are predominant forces, for better or worse.'' \citep[this issue]{Holton2014}

The themes of science in human culture, ethics, and the impact of science on society and individuals, as well as the felt responsibility to students, was echoed in the remarks of all of the subsequent speakers.

Michael Matthews reminded the conferees that the history of the use of the history of science to teach science extends at least back to the beginning of the twentieth century. In multiple public reports, and in texts on science teaching, the history of science has been viewed as the means for introducing to students the vitality and philosophy of scientific inquiry. However, despite projects such as Project Physics, and Leo Klopfer's History of Science Cases \citep{Klopfer1963}, methods for communicating a philosophical understanding of science, and using the history of science to encourage greater student engagement, have not survived the pressures of classroom content instruction and teacher preparation. Matthews made this ``melancholic observation'' with a citation from Klopfer, ``[p]roposals for weaving the history and nature of science into the teaching of science in schools and colleges have a history of more than sixty years. (\ldots) These ideas anchored a web, and the strands of science content and science history formed the web's pattern. Yet each of these webs was fragile; they rarely persisted for very long and left little trace on the science education landscape.'' \citep[105]{Klopfer1992}

In his response to Holton and Matthews, Sevan Terzian expanded on the need to embed science education in the broader landscape of the humanities and social sciences. In the process, Terzian provided a historical and political perspective on science education policy with its current emphasis on preparing more students for science, technology, engineering, and mathematics (STEM) careers. This he said was similar to the vocational focus of the high schools at the turn of the twentieth century, and is now becoming common in higher education. What suffers when vocationalism is emphasized are the humanities and social sciences, and more generally the liberal arts education that ``fosters critical thinking, creative problem solving, and empathy.'' These are qualities necessary to prepare students for participatory democracy, an objective that suffers in the absence of science education in a liberal arts context through which ``broader segments of the public gain not only some understanding of science but an appreciation for how it works.''

The themes that emerged in the subsequent talks were framed by Gregory Kelly's presentation titled ``Philosophy of Science and Science Education Reform.'' Kelly framed the issues along lines similar to the initial motivating questions for the conference:

\begin{list}{-}{}
\item Learning: ``How can the philosophy and history of science contribute to learning opportunities that incorporate critical discourse?''
\item Science curriculum: ``How can philosophy and history of science contribute to curricula that supports citizens' abilities to decipher, analyze, and participate in socioscientific issues?''
\item Science teacher education: ``How can philosophy and history of science develop in teachers a critical stance toward science, views of science, and science and engineering standards and curricula?''
\end{list}

The issue of critical discourse and analysis, and its connection to philosophy, may have been the most significant thread that wove through the conference presentations. As expressed by Melissa Jacquart, one of the graduate students invited to attend from Western University: ``Improve students' critical thinking skills, and it can improve their understanding of science'' \citep{Jacquart2013} and by extension improve their understanding of socioscientific issues.

As discussed by Richard Duschl, the \emph{Next Generation Science Standards} (\emph{NGSS}) \citep{Achieve2011} integrates the teaching and learning of the nature of science into the expected outcomes. The \emph{NGSS} relies upon the \emph{Framework} for an organizing rubric that specifies eight classroom practices for the teaching science of science and engineering, fundamental cross-cutting concepts that span the sciences, and specific core content for each of the disciplines (life sciences, physical sciences, earth and space sciences, and engineering, technology and applications of science). Through such science practice with specified content objectives, students are to be supported in achieving the four strands of science proficiency encouraged by the National Research Council \citep{NRC2008}: (1) understanding scientific explanations; (2) generating scientific evidence; (3) reflecting on scientific knowledge; and, (4) participating productively in science.

In support of the \emph{Framework} and \emph{NGSS}, Duschl cited the naturalistic philosophy of science that has evolved from efforts to understand the practice of science by scientists \citep{Duschl2013}. The learning goals corresponding to this philosophy are a combination of conceptual, epistemic, and social activities \citep{Duschl2008}. The outcome is a ``new view of NOS,'' from which follow new performance expectations for teaching the nature of science. The new view emphasizes the role of models and evidence in theory development, ``sees the scientific community, and not individual scientists, as an essential part of the scientific process,'' and views the scientific enterprise as relying upon a variety of tools such as ``instruments, forms of representation, and agreed upon systems for communication and argument.'' \citep{Duschl2013}

In his response to the presentations of Duschl and Kelley, Jacob Foster, State Director of Science and Technology for the Massachusetts Department of Education gave a \emph{realpolitik} view of how standards are interpreted at a state level. Standards specify to a state's education community the expected outcomes, not the methods to be used to accomplish the outcomes. Thus, for nature of science outcomes specified in the \emph{NGSS}, ``history and philosophy of science may be one excellent way to engage them [students] in getting there [outcomes] but so are many of the other ways that we know in science education.'' Foster also reminded the audience that ``science and mathematics standards for that matter are not about preparing the select few for STEM majors and careers; they are the floor for all students, what we expect for all students in terms of what we expect for being scientifically literate.'' Herein lies an example of the slipperiness of what scientific literacy is. If the school perspective is that scientific literacy has been achieved when student assessment shows that the desired outcomes of the standards are satisfied, then without explicit outcomes in the standards that students should be able to understand how to apply their understanding of science to their personal lives, and how to apply their scientific knowledge to matters of public policy, the state view of scientific literacy and what the HPS community expects from a scientifically literate student, will deviate substantially.

In his presentation, Fouad Abd-El-Khalick reviewed the use of the history of science for teaching the NOS, and the positive outcomes of the studies of curricula that relied on case studies and the history of science (e.g., Project Physics \citep{Holton1967,Holton2003} and Klopfer's History of Science Case Studies \citep{Klopfer1963}). In teaching about the NOS, Abd-El- Khalick refers to history of science (HOS) as the ``stuff'' of NOS, the material for critical analysis, for otherwise, ``[w]ithout HOS, teaching about NOS could be reduced into a set of de-contextualized platitudes.'' To provide the context, interventions should combine ``rich historical case studies'' along with scientific internships and inquiry-based contexts. In each case, reflection on the case study or activity in the context of nature of science attributes is essential. Such reflection is consistent with the Strands of Scientific Proficiency \citep{NRC2008} and is a research based conclusion \citep{Abd-El-Khalick2013}.

Abd-El-Khalick also sounded a pessimistic view about the likelihood of broadening science education to include the history of science. He identified one source of this problem with teacher preparation programs. Abd-El-Khalick observed that many of these programs leave little room for a science teacher to elect history of science courses. Moreover, even if the science teacher does take such a course, ``the mounting pressures on schools associated with frequent and high-stakes testing continues to push against goals, such as enabling students to develop more sophisticated understandings of how knowledge is produced and justified in the sciences, in favor of developing speed and accuracy in applying scientific formulæ to solve word problems and get the right answer.''

Addressing the affective side of using HPS for science education, Fanny Seroglou offered ways to engage a wide age range of students in the history of science so as to foster in them a better grasp of the nature of science. To ``give meaning to science and connect science to culture and society'' she uses multimedia, especially film, in a teacher training course on science and culture. Based on the response of these students, Seroglou concludes that non-science pre-service and in-service teachers ``appreciate science teaching with a strong cultural perspective as it offers them motive and inspiration to learn and teach science.'' Outcomes from this approach show measurably improved understanding of the consensus components of the nature of science. In other work, Seroglou has also demonstrated the value of engaging younger children through the history of science. Her group has worked with 12-year-olds and engaged them interactively by having them develop their own animated movies about the history of astronomy \citep{Piliouras2011}, and used the development of stories about science and scientists when working with 7-year olds.

The question of how to productively engage students in learning science \emph{content} through the history of science, as opposed to the nature of science, was addressed by Xiaodong Lin-Siegler. The traditional sidebar in textbooks provides biographical information about scientists. But what aspects of a scientist's life appeal most to students and assist them in learning science? A physics teacher inclined to refer to Einstein's role in the history of physics is likely to cite his annus mirabilis in which he presented solutions to multiple major puzzles in physics, and not dwell on Einstein's personal struggles to achieve academic success, or overcome subsequent prejudice against his methods and ethnicity. Lin-Siegler's research \citep{Hong2012} provides evidence that students are more likely to be motivated to understand and retain science content if the historical account dwells more on the scientist's struggles than successes.

Douglas Allchin's talk bridged the presentations about motivating students through the use of the history of science, or sociocultural connections with science, with the talks that followed that dwelled on how to construct a socioscientific curriculum, the cognitive challenges inherent for students learning such material, and the barriers encountered by teachers. Allchin reminded the audience that the reason for our interest in students understanding the nature of science is to prepare them to be members of a democratic society who will have to make decisions on a wide range of socioscientific issues that affect them directly and indirectly; these issues include: advice by medical panels, and environmental issues such as climate change and hydraulic fracking. These socioscientific issues are multifaceted and require a broad understanding of the scientific enterprise within society, including providing students with institutional knowledge of who is an expert, whom you can trust, where the funding comes from, and what the scientists' biases and conflicts of interest may be. Throughout his talk, Allchin referred to the relevance of the contributions of contemporary philosophers and historians of science to our understanding of socioscientific issues (e.g., who is an expert), the nature of inquiry, and the value of the learning community.

While advocating for this broader education in the nature of science, Allchin emphasized the need to be practical with respect to the needs of teachers, and the response of students in the classroom. Allchin made the points that, for student motivation, the story is the message, not the specific content; that NOS cannot be ``funneled'' into students, but must be taught explicitly and with active, inquiry-based learning and problem-solving; and, that teaching about the nature of science and socioscientific issues cannot be yet another add-on in the classroom \citep{Allchin2014}.

In his response to the presentations, Frank Keil raised several categories of research questions that need to be resolved about the use of the history of science for science education. He asked what version of history will be taught (e.g., achievements in chronological order? historical evolution of ideas? individual stories?), and what are the motivational issues (e.g., scientists more human? de-motivation since not all struggles are perceived as good? different effects at different ages?). He raised similar lists of questions for cognitive factors in the use of the history of science for science instruction. Keil concluded by observing that ``[e]mbedding science content in history is [a] multidimensional process with a wide array of potential effects'' and ``[b]oth Motivational and Cognitive effects could be either positive or negative.''

Within this context of teaching and learning about the nature of science, whether from the perspective of inquiry or from the perspective of socioscientific issues, Michael Clough emphasized the importance of the metacognitive ``why'' for students and teachers. Students need to understand \emph{why} they should study the nature of science, just as they need to understand \emph{why} scientists seek naturalistic explanations, and why idealization is useful. The same questions must be answered for teachers. To be effective, teachers must have a deep enough understanding of the nature of science themselves, ``[t]he nature of science is everywhere and so they (teachers) can pull it out in the context they find themselves [in].'' To provide students with the necessary scaffolding, teachers must be prepared to draw students' attention to the nature of science at teachable moments, and it is not easy prepare teachers to do this. ``Effective history and nature of science instruction demands competent teachers. And teachers who understand the history and nature of science understand science content, understand how to weave it together, and know how to teach well.''

The linkage between socioscientific issues and the nature of science is either implicit or explicit in many of the presentations reviewed above. Keil's researchable concerns over how best to use socioscientific issues to motivate students, Seroglou's success in using them to motivate student learning, Allchin's compelling reasons for why students need to understand the nature of science to understand how to address socioscientific issues, and Clough's concern over how to prepare teachers all weave together and beg for an answer as to whether a curriculum can be designed that can address science and socioscientific issues successfully.

In this context, the \emph{Exploring Bioethics} \citep{EDC2009} curriculum discussed by Mildred Solomon is a curriculum case study itself. The objectives of the developers were for students to ``develop the ability to critically reason in the moral domain'' so as to equip them ``to handle the many ethical challenges they will face.'' Students need to be prepared ``with the skills and critical tools to make well-informed personal decisions and participate thoughtfully in forging public policies about ethical issues related to advances in the life sciences.'' While this curriculum focuses on the life sciences, its approach of distinguishing between scientific, legal, and ethical questions is a model for curricula that provides students with the opportunity to critically reason about issues such as the use of science for the national defense, the invasion of privacy by technology, and anthropogenic environmental impact. In the evaluation study that Jeanne Chowning presented on the effect of the \emph{Exploring Bioethics} curriculum, the evidence points to students being motivated through the discussion of ethical issues, understanding the connection between science and society, and engaging in critical thinking and analysis.

Tina Grotzer's talk explored the underlying cognitive demands that must be addressed in science education when the student is confronted with challenges that mix evidence, perception, and experience. Her specific research focuses on a computer simulation designed to help students disentangle causative variables and distinguish their effects. However, the argument Grotzer offered of how we must account for students' perceptions, determined both physically and emotively, and their interaction ``with the nature of science and ultimately, public understanding of scientific research'' is a theme that cuts across all attempts to use case histories, either historical or simulated, to develop students' judgment for socioscientific issues.

In her response to Solomon's and Grotzer's presentations, Luciana Garbayo observed that the understanding of scientific models and their applications are crucial for proper moral reasoning in bioethics, adding layers of complexity to their teaching and learning. She suggested that the engagement in bioethics reasoning requires both a modicum of understanding of ethics and of epistemology to address the gamut of epistemic disagreements with moral relevance in a scientific context \citep{Garbayo2014}. In the context of Solomon's and Grotzer's work, the outcome for teaching and learning is that there is a greater demand on students' cognitive and emotional resources in aligning HPS contexts with bioethical reasoning, so that epistemic and moral disagreements are properly understood. Garbayo's observations about the additional complexity associated with bioethical reasoning clearly generalize to any curriculum that might treat socioscientific issues where science and morality meet.

The tension in science education about using HPS to teach critical reasoning skills in the science classrooms emerges when reviewing the obstacles perceived by teachers and science educators engaged in teacher preparation and professional development. In the presentation of the results of his research, H\"ottecke followed up on earlier work \citep{Hottecke2011} and discussed the multiple barriers for science teachers to including HPS in the classroom. While his studies were conducted with physics teachers, there is little reason to believe that his conclusions do not apply to science teachers generally at the high school level. The obstacles include ``lack of an effective implementation strategy for HPS,'' the ``culture of teaching physics,'' ``physics teachers' attitudes and beliefs'' and ``institutional boundaries'' similar to those referred to by Abd-El-Khalick. In addition, science teachers are uncomfortable with the aspects of science and reasoning that are highlighted by a review of the history of science, and are essential to an understanding of the NOS. Uncertainty and ambiguity, two aspects of authentic science research, and which lead to the struggles which culminate in new science, are difficult for science teachers to acknowledge in the science classroom.

Despite the limitations on teacher preparation programs cited by Abd-El-Khalick, some do include the history of science to prepare student teachers in NOS. David Rudge provided an example of the use of a case history that helped pre-service teachers better understand NOS. Specifically, the historical case referred to the mutation of the coloration of a moth which proved to be, from the perspective of natural selection, an adaptive mutation. The students who took this course showed gains in their understanding of the NOS, most specifically the nature of theories. The instruction for this activity included time for discussion and reflection by the students \citep[this issue]{Rudge2014}.

A means for preparing teachers to understand how scientific models are developed was offered by John Clement. Models are at the core of the \emph{Framework}'s practices and any understanding of the nature of science. Clement has studied how experts develop models to solve problems. Based on his research, he has created a schema for the model formation process. He has found that this schema also fits historical examples of model development. The process represented by the schema relies on a cycle of generating, evaluating, and modification of a model \citep{Clement2009}. Clement proposed that by introducing teachers to this schema it may be possible to help them better understand how to introduce models into their own classrooms and allow their students to engage in modeling activities. This would be a big step for moving classroom instruction in the direction of the new science practices described by the \emph{Framework}.

Barbara Crawford provided an example of an immersive experience in scientific research on teachers' understanding of the nature of science, and the subsequent impact on their students' understanding. The teachers were immersed in inquiry through a paleontological investigation of the strata of an ancient seabed. During this investigation they studied the science content and reflected about the nature of science. Afterwards, the teachers engaged their own students in inquiry exercises about fossils. Crawford found that the immersive research experience resulted in significant gains in the teachers' understanding of the nature of science, and their students' understanding as well.

Two of the plenaries dealt directly with science and history. Bob Bain's presentation was on \emph{Big History} (\href{https://www.bighistoryproject.com/portal}{https://www.bighistoryproject.com/portal}), an approach to history on the grand scale of deep time. The objective is to help a person/student understand the physical development of the universe, and where humanity fits into this history. It is a grand (and scientific) origins story. From the perspective of HPS, this project is at the intersection of history, philosophy, and science. The \emph{Big History} project represents another way to engage students with science through a history based on science, and one that raises the deepest existential philosophical questions. The triad of perspectives represented through history, philosophy, and science offer the opportunity for teaching and learning about the demarcations between the three.

By comparison, the plenary address by Patrick Morris focused on the history of just one person, Sir Isaac Newton, as dramatized in the play \emph{Let Newton Be!} performed by the Menagerie (\href{http://www.menagerie.uk.com/productions/archived/newton/}{http://www.menagerie.uk.com/productions/archived/newton/}) \citep{Stott2009}. The playwright, Craig Baxter, and the director, Patrick Morris, offer insight into the efforts of a singular genius who revolutionized science. While a ``great person'' approach to the history of science may not lend itself well to an understanding of the nature of science, neither can history ignore seminal contributions. Discussion of such a theatrical production lends itself better to an English or social studies class than to the science classroom, and provides an example of how science as part of human culture can be appreciated by students.

After the conference, the four graduate students from Western University posted texts on the blog of the Rotman Institute of Philosophy (\href{http://www.rotman.uwo.ca/2013/02/}{http://www.rotman.uwo.ca/2013/02/}) that highlighted central points discussed during the conference. The short articles of this blog captured the principal themes emphasized by the speakers and that subsequently threaded through the workshops. Here we use the titles of the blogs as summary of the topics of the above presentations and as preface for the next section.

\begin{enumerate}
\item Improving Scientific Literacy Through Improved Critical Thinking Skills \citep{Jacquart2013}
\item How to include the history and philosophy of science (HPS) in science education
standards? \citep{BenetreauDupin2013}
\item Broadening the Goals of Science Education –- ``how HPS might be used to improve
science education in the areas of: ethics of science, the role of science in a democratic
society, and the evaluation of scientific evidence.'' \citep{Brandt2013}
\item What must be done to educate, equip, and support teachers to incorporate HPS into their
curricula? \citep{Fox2013}
\end{enumerate}

\section{Working Group Conference Reports}
\label{Groups}

On the second day, the conferees broke up into working groups to discuss and report on research issues related to the use of HPS for science education. Five working group topics emerged from a set of initial suggestions from the conference organizers, and then modified by input from the conferees. Conferees were asked to join one of the working groups at the beginning of the second day and to prepare a draft report by the end of the afternoon. Working group chairs remained for a third day to work on drafts of their reports before departing. Their charge included elaborating, in the spirit of the working group's discussion, on the notes taken during the working group meetings to provide a report for the conference organizers.

The reports are uneven in their detail corresponding to the difficulty of the topics, and the need for a group of experts from different fields to express their views in a short period of time. What follows is a digest of the four working group reports that arrived at conclusions.

In order to better understand the role of HPS in preparing the student as a citizen in modern society, the working group chaired by Douglas Allchin (University of Minnesota) addressed the question of \emph{How can history and philosophy of science (HPS) contribute to students' and pre-service teachers' understanding of the nature of science, in the context of a scientifically literate citizenship?} Reflecting its topic, this working group called itself the \emph{NOS and SSI Working Group} where SSI is an acronym for socioscientific issues. This working group dissected its question thoroughly and generated many research questions.\footnote{For the full report see \href{http://www.bu.edu/hps-scied/files/2014/04/WG1-HPS-NOS-and-Scientific-Literacy.pdf}{http://www.bu.edu/hps-scied/files/2014/04/WG1-HPS-NOS-and-Scientific-Literacy.pdf}.} Only a few of the questions posed are reviewed here.

Because an initial understanding of NOS is fundamental for students to engage in discussion about socioscientific issues that confront citizens, the working group first addressed student learning of NOS. Given that different researchers have different definitions of NOS, the group took a broad interpretation of NOS, but specifically included an understanding of the enterprise of science necessary for making informed societal decisions. This broader perspective relies on contemporary science studies \citep[this issue]{Allchin2014}. While acknowledging the existence of many studies, the working group identified specific questions in need of research. The first dealt with the question of the formats for students to learn about NOS. The group asked whether the approach should be student-based investigatory experiences, historical cases, contemporary cases, or research apprenticeships? For each of these cases, how should the teacher be structuring the lessons? The group further identified the development of assessments of students' learning of NOS as applied to socioscientific issues as a challenge. As during the conference, the question of how to teach non-experts to engage in critical reasoning founded on evidence, and what skills a teacher must have for this purpose, was asked.

The group chaired by Christine Baron (Boston University) tackled the question of how to prepare students (K-2, 3-5, 6-8, 9-12, 13-16) or pre-service teachers to reason about issues raised by science which cannot be addressed within the domain of science alone, for example, ethical questions, economic questions, and cultural issues. Based on the analysis that was the outcome of this group, we will refer to this working group as the Critical Reasoning and Discourse (CRD) Working Group.

The need for developing critical reasoning skills for students across the disciplines and the need for the professional development of teachers to support critical reasoning through discourse dominate the recommendations of the CRD Working Group:

\begin{list}{-}{}
\item ``For students to have a thorough understanding of science, it is essential to include issues of the role the ethics in science, the evaluation of evidence, and the role of science in a democratic society into K-12 classrooms. However, the discussion of these issues cannot and should not be limited to the conversations students have in their science classes. It is critical that we draw support from teachers in other disciplines, particularly social studies teachers, to help facilitate these conversations.''

\item ``[I]t is clear that content training alone will not provide sufficient support for either science or social studies teacher to effectively facilitate discussion of these issues.''
\end{list}

Citing the eight research practices of the \emph{Framework for Science Education K-12}, this working group concluded that: ``[t]hese practices are not unique to the natural or physical sciences. They are also similar to those cited in the Common Core State Standards in Mathematics, English Language Arts, and other fields as well.''

The CRD Working Group report went on to say that 

\begin{quote}
many teachers, in any subject area, are not skilled in orchestrating the kind of reasoning-based discussion required for discussing complex issues in science or society. (\ldots) [T]he productivity of classroom discussion for building the targeted capacities requires skilled pedagogy. (\ldots) Thus, we strongly recommend the development of a framework that will include goals for discussion, planning for discussable questions, building norms for respectful and equitable discussion, and a series of ‘talk tools' that teachers may use to learn to orchestrate skillful discussion moment to moment with a large group of students. 
\end{quote}

\noindent In many ways, the CRD Working Group is calling for the kind of detailed curriculum constructed for Exploring Bioethics. 

The Teacher Preparation Working Group chaired by David Rudge (Western Michigan University) focused principally on designing a way to produce and disseminate materials for science teaching that use HPS materials. The group observed ``that while some HPS lesson plans and modules exist, they are not often available to teachers, either because they are published in scholarly journals for which the teachers have no access, or because they are provided in a form that is not user friendly.'' To remedy the accessibility problems, the Teacher Preparation Working Group proposed the establishment of an online portal to collect and disseminate materials. To make the portal more germane to teachers' needs, the materials would include references to the \emph{NGSS}, notes by teachers who have used the materials, and access to a local expert on HPS. Finally, to ensure the usability of these materials, the lesson plans would be created at summer institutes that brought together HPS experts and high school teachers. These lesson plans with supporting material would then be posted on the portal.

The Policy Working Group chaired by Michael Marder (University of Texas Austin) began its report by remarking that many of the conferees chose to participate in this group because one of the subquestions dealt with the equity in education. The group felt that in service of providing education to all student groups, the ``skillful use of the history, philosophy, and social context of science can play an important role in teaching that motivates and empowers students learning science.''

The Policy Working Group began by generating questions that might warrant NSF funding. These were general questions:

\begin{list}{-}{}
\item Is introduction of history and philosophy of science into science instruction most effective for future scientists, future educated citizens, or both?
\item What are the practical ways for teachers to manage this introduction that are consistent with time and resource constraints?
\item Is it possible to determine in large-scale well-controlled studies whether introduction of history and philosophy of science into science instruction has beneficial effects on participation and other desired outcomes for under-represented groups?
\item How can the history and philosophy of science illuminate the ways that science is part of society, and society shapes research?
\end{list}

Having posed these questions, the Policy Working Group suggested that the National Research Council (NRC) should sponsor a report on what is already known from prior NSF and other Federally funded research on the use of HPS for teaching NOS and science literacy. 

For more immediate impact, the Policy Working Group identified the adoption of the Common Core and the \emph{NGSS} as a singular opportunity to infuse new curricular materials into the classroom. Whether or not this opportunity materializes, the suggestion reflects the more general lack of HPS based science curriculum observed by the Teacher Preparation Working Group, as well other working groups and during the public conference talks.

In addition to the working groups, some of the philosophers in attendance (Alisa Bokulich, Katherine Brading, Carol Cleland, Patrick Forber, Luciana Garbayo, and Christopher Lehrich) congregated ``to construct ways to collaborate and contribute to discussions on the uses of HPS in science education.'' As a group, they noted the discontinuity among the HPS communities and science educators in the use of many key concepts central to understanding science. This discontinuity, they argued, may stem from the fact that an extensive contemporary philosophy of science literature has yet to be incorporated into discussions about NOS. Contemporary work in the philosophy of science and science studies has moved well beyond falsificationist and Kuhnian accounts of science, and has become more sensible to what constitute science as a practice and its sociocultural environment \citep[see][this issue, for a review of such recent developments]{Allchin2014}.

\section{Discussion}
\label{Discussion}

In reviewing the conference as a whole, the presentations of the speakers and the products of the working groups, common themes emerge. 

\begin{list}{-}{}
\item \textbf{Students need to understand that central to science is argumentation, criticism, and analysis.} As expressed by the conferees in their presentations and reports, central to the nature of science is argument based on critical reasoning that cites evidence and/or well-supported theory. Michael Ford expressed this when he remarked that ``science is really about arguing about data'' and ``you provide evidence for a reason. What's the reason? Because you are trying to convince somebody else. (\ldots) [T]he reason that you bring evidence is because somebody else is healthily and rationally skeptical.''

\item \textbf{Students should be educated to appreciate science as part of our culture.} This emerged in the talks of Gerald Holton in emphasizing the place of science in a liberal arts education; in the appeal by Sevan Terzian that science education not be separated from the humanities and the social sciences in the name of STEM competitiveness; in the themes listed by Gregory Kelly; in the remarks of Katherine Brading on recognizing science as part of our culture as we do art and music; and, in the use of sociocultural connections to science to motivate students by Fanny Seroglou.

\item \textbf{Students should be science literate.} Here we interpret science literacy as the preparation of students to be ready to make decisions when science affects them personally, and to make decisions as citizens about science policy and contribute to a democratic society. The importance of science literacy was emphasized in the presentations of Douglas Allchin and Mildred Solomon on socioscientific issues; by Sevan Terzian in his emphasis on preparing students for participatory democracy; and, in the reports of the working groups on NOS and SSI and on Critical Reasoning and Discourse.

\item \textbf{What is meant by the nature of science as discussed in much of the science education literature must be broadened to accommodate a science literacy that includes preparation for socioscientific issues.} This was a conclusion of the NOS and SSI working group, and implicit in the reports of the Critical Reasoning and Discourse and the philosophers' working groups.

\item \textbf{Teaching for science literacy requires the development of new assessment tools.} The need for new assessments related to scientific literacy was explicit in the NOS and SSI report, but also implicit in other presentations and reports, since assessment ultimately is necessary for the teaching linked to the changes in pedagogy recommended by the Critical Reasoning and Discourse working group and the Policy working group. 

\item \textbf{It is difficult to change what science teachers do in their classrooms.} This was emphasized by Abd-El-Khalick, Allchin, Clough, and Höttecke. The demands on science teachers are stringent. High stakes testing that stresses content mastery, the time constraints of class length, and the wide range of teachers' collateral responsibilities during the work day, all militate against introduction of new materials into the classroom that may appear tangential to science content mastery. 
\end{list}

There is an unaddressed conflict in the outcomes of the presentations listed above with current national and state science standards. On the one hand, the nature of science is considered part of science education, with categories of its nature along with learning outcomes specified in the \emph{Next Generation Science Standards} \citep{Achieve2011}. On the other hand, the speakers in general expressed the view that one of the most important educational objectives for a liberal arts education should be the preparation of students to be citizens in a participatory democracy. This requires the linkage between science education and education in the humanities and social sciences. 

The learning outcomes of the \emph{NGSS} do not connect with this latter objective. A review of the \emph{NGSS} shows that the nature of science outcomes are intended to be intrinsic to science alone and discussed in a way that limits connections to science policy. The effort to decouple the learning outcomes from connections to policy is exemplified by the standard MS-ESS3-4 which connects to the Nature of Science category of Science Addresses Questions About the Natural and Material World. The standard reads ``Scientific knowledge can describe the consequences of actions but does not necessarily prescribe the decisions that society takes.'' This can be understood as an attempt to distance science instruction from all of the thorny policy and ethical questions that arise from the teaching and learning of science content, even when NOS learning is an expected outcome.

Similarly, at a time when there is an enormous need to educate citizens about the sources of legitimacy and authority in science so as to silence the counternarratives manufactured by industry and fringe political groups \citep{Oreskes2010}, the NGSS Appendix H suggests the use of politically neutral history of science examples such as the ``Copernican Revolution, Newtonian Mechanics, Lyell's Study of Patterns of Rocks and Fossils'' and so forth.

This schism most likely reflects the political realities of the United States where curricular decisions are made at the State and regional or local school board level. The National Science Foundation has already experienced the budgetary wrath of Congress when it was perceived to be developing a national curriculum in the sciences \citep{NRC2007}, much less a curriculum that would encourage students to discuss socioscientific issues that have connections to ethics and religion.

There is a second reality that obstructs the use of the science classroom in the preparation of citizens who are scientifically literate in the sense that \citet{Allchin2013} writes ``Students should develop a broad understanding of how science works to interpret the reliability of scientific claims in personal and public decision making'' (p. 4). As Abd-El-Khalick, Allchin, Clough, and H\"ottecke made clear in their presentations, there are multiple barriers for teachers to adopt the history of science in their classrooms to teach the nature of science. The barriers include the lack of curriculum and strategy for use of history for teaching science; the culture, preparation, and attitudes and beliefs of science teachers; and, institutional barriers which in the United States include the heavy emphasis on standardized tests that do not include history of science content. That these barriers are high is empirically confirmed by the history of the attempts to include HPS into science education. Moreover, to educate students to be scientifically literate will require an even more extensive curricular effort with teacher professional development.

Abd-El-Khalick makes the point that history of science is the ``stuff'' of the nature of science. The real problem may be that the HPS community is not focusing on the right stuff as history if the objective is to prepare students with scientific literacy for citizenship. All the speakers concur in the need to develop students as critical thinkers and that this is essential to prepare citizens to evaluate socioscientific issues. However, expecting students to make the transfer of an understanding of the nature of science, as assessed using the \emph{NGSS} learning outcomes, to making reasoned decisions about personal needs or public policy defies educational experience. Moreover, while including case studies from the developmental history of science may be effective in improving students' attitudes towards science and, if coupled with reflective exercises, improve their understanding of the nature of science, on the whole such examples from history do not prepare students to reason about the impact of science on society.

The right history stuff is what motivates students (and teachers) and connects explicitly to issues of science in citizenship. This is the history of socioscientific issues, the issues where science is a player in the political arena. Many examples are laid out in \emph{Merchants of Doubt} \citep{Oreskes2010}. The consequences of medical knowledge about smoking, cancer, and stem cell research; issues related to national defense such as nuclear weaponry, smart bombs, drones, the strategic defense initiative; and, and environmental issues including acid rain, the ozone hole and global warming, are all science issues for which citizens must decide policy. It is bad enough that evolution theory denial obscures for the citizenry the threat of the abuse of antibiotics and the irrelevance of DDT for malaria control \citep{Oreskes2010}. But, climate change denial as governmental policy has changed the urgency for preparing students to understand science, the scientific enterprise, and the connections between science and social and economic policies. What we live now is history too, and it is history we create and understand through natural science. The consequences of climate change denial will be central to education in the humanities and social science by the end of the twenty-first century, and for many centuries to come. For educators, it will be part of the history of the failure of science education for the common good.

The \emph{Exploring Bioethics} curriculum described by Mildred Solomon is a step in the right direction in preparing students through critical thinking to evaluate medical and biological scientific advances both in their personal lives and in their public participation as citizens. While this curriculum has been tested in biology classrooms with positive outcomes, its counterparts in chemistry and physics are not available and, for the reasons already given, it is unrealistic to expect similar curricula in the physical sciences to be widely adopted. Moreover, given the observation by Matthews of the melancholic history of incorporating HPS into the science curriculum, even such an excellent curriculum as \emph{Exploring Bioethics} is likely to fade from classrooms.

If there is a breath of hope for the liberal arts education necessary to prepare students to be scientifically literate and capable citizens in an age dominated by scientific advances, it was voiced in the report by Christine Baron for the Critical Reasoning and Discourse Working Group. This report observed that ``for students to have a thorough understanding of science, it is essential to include issues of the role of ethics in science, the evaluation of evidence, and the role of science in a democratic society into K-12 classrooms. However, the discussion of these issues cannot and should not be limited to the conversations students have in their science classes. It is critical that we draw support from teachers in other disciplines, particularly social studies teachers, to help facilitate these conversations.''

In the \emph{Framework}, the Three Spheres of Scientific Activity succinctly summarize the practices of science. The central sphere is comprised of ``Argue, Critique, Analyze,'', the activities of critical thinking. These descriptors for critical thinking apply across the spectrum of academic effort, whether it is the humanities, social sciences, or the sciences. As remarked by Katherine McMillan, in her response to the panel with Richard Duschl and Gregory Kelly, when confronted with history teachers she recognized that the principles of argumentation are common to instruction in history and science. Similarly, the CRD working group observed that the eight practices of the \emph{Framework} are ``similar to those cited in the Common Core Standards in Mathematics, English Language Arts, and other fields as well.''

Since the time that James \citet{Conant1957} recognized the need to educate college students about science through historical examples, the history of science has grown so that the ``stuff'' that students now need to learn for their citizenship extends beyond the science classroom to the social studies/history classrooms, and even beyond, as Seroglou has shown, to the humanities/ literature/ arts classrooms. To educate scientifically literate students, the community of science educators (especially those interested in advocating an HPS-informed approach) needs to enlist social studies and humanities teachers to extend the courses over which the enterprise of science is taught. This does not reduce the need for science content education. Indeed, the disciplines that may need to change their curriculum the most are the humanities and social studies. Such a conversation among educators is necessary in order to create a curriculum that educates our students to be critical thinkers and make sense of a world in which ``science and technology are predominant forces'' as Holton remarked at the conference \citep[this issue]{Holton2014}. 

As Katherine Brading said in response to the panel with Allchin and Seroglou, ``what is really important is giving all of our students and all of our children the sense that they have ownership over science, that they have the authority to speak about science and have a voice that expresses that ownership. To give them a sense of their cultural inheritance, that science is part of their cultural inheritance.'' This plea for students to receive a humanistic education in science, and for science educators to recognize that this is necessary to be successful in their own mandate, brings us full circle to Holton's exhortation for science as one of the liberal arts.

\section{Conclusions}
\label{Conclusions}

The objectives of the conference specified outcomes that included a research agenda related to curriculum, student learning, and teacher development. The working group reports provided a range of directions for future research. The broad set of researchable questions generated by the NOS and SSI Working Group that surround preparing scientifically literate students was summarized above, and their complete report is posted on our website; the researchable questions of the Policy Working Group were also reviewed above. The Teacher Preparation Working Group's focused on making HPS based materials accessible to teachers, although what the teachers did with such materials would surely be researchable.

The Critical Reasoning and Discourse Working Group's report dealt directly with the question of attempting to clarify the role of HPS in the general K-16 curriculum. With the understanding that science literacy refers to preparing students as citizens to make decisions relating to the impact of science on their personal and social lives \citep{AAAS1990}, their conclusion is that teaching for science literacy straddles the content domains of history/social studies, humanities and science, and that the reasoning skills to be taught in these different domains largely overlap.

Inclusion of the history and philosophy of science in science education has historically had two rationales. One is that HPS can be used to improve science content learning and the preparation of scientists. The second is that an understanding of HPS is necessary to prepare students to be scientifically literate, a preparation necessary for modern citizenship. With these two objectives in mind, the evidence and advisement provided in the conference presentations and the working group reports support the following conclusions.

\begin{list}{-}{}
\item In order to prepare students to be citizens in our participatory democracy, science education must be embedded in a liberal arts education. Science literacy requires science content learning, including NOS, along with reasoning skills learned in social studies and developed through arguments about ethics.

\item On their own, science teachers cannot be expected to prepare students to be scientifically literate. Science literacy requires learning that extends beyond the science classroom into the social studies and humanities classrooms. The barriers for science teachers to include HPS for science literacy are well established and historically supported.

\item To educate students for scientific literacy will require a new curriculum that is coordinated across the humanities, history/social studies, and science classrooms. In our fast moving age, the history of science has accumulated a new history over the past seventy-five years, one of socioscientific impact that affects all citizens. This history needs to be brought into the K-16 curriculum to provide the ``stuff'' for students to study. It is likely that studying this recent and relevant science history will be motivating for students. Based on the outcomes of prior efforts to embed science in a liberal arts education, there should be an improvement in attitudes towards science. Coordinating science content instruction with science history taught in social studies and humanities classrooms is likely to improve science content retention as well.

\item The current science standards are improved with respect to outcome expectations for the nature of science in relation to science. However, these outcomes do not include explicit learning outcomes for students to evaluate the impact of science on their personal lives (e.g., through medicine, technology, and consumer products) and science as it affects public policy (e.g., national defense, privacy, health, the environment).
\end{list}

There are political reasons within the United States to expect resistance to efforts to promote new curriculum that prepares students to be scientifically literate citizens. However, the empirical evidence that requires this effort is before us now. Science denial is an accepted political strategy and its damaging impact on the environment surrounds us. This science denial extends to the rejection of well-reasoned arguments by experts outside the natural sciences such as economics. The only resolution to this problem is a new curriculum that spans K-16 classrooms and is designed with history for evidence, science for explanation, and philosophy for analysis.

\vspace{2em}
\subsection*{Acknowledgements}

\small{The authors gratefully acknowledge the assistance of Emily Allen and Thomas Hunt, Doctoral Fellows in the School of Education at Boston University, and Dr. Charles Winrich, Director of Science at Babson College, for their readings of drafts of this manuscript.

For the logistics of the conference organization itself, we are greatly indebted to Thomas Hunt. It was his organizational skills that allowed us to maintain a database of all the invitees and eventual conferees. Mr. Hunt further designed and created the conference website, organized the video recording, obtained the permissions, and uploaded the conference video and PowerPoint presentations.

The conference was funded by a conference grant from the United States National Science Foundation (NSF) from the Division of Research and Learning, (REESE-1205273). We are grateful to the program officers of NSF for their recommendations of conferees.}

\bibliographystyle{chicago}
\bibliography{/Users/yannbenetreau/Library/texmf/bibtex/bib/library}

\begin{thebibliography}{}

\bibitem[\protect\citeauthoryear{AAAS}{AAAS}{1990}]{AAAS1990}
AAAS (1990).
\newblock {\em {Science for All Americans (Project 2061)}}.
\newblock New York: Oxford University Press.

\bibitem[\protect\citeauthoryear{Abd-El-Khalick}{Abd-El-Khalick}{2013}]{Abd-El-Khalick2013}
Abd-El-Khalick, F. (2013).
\newblock {Teaching with and about nature of science, and science teacher
  knowledge domains}.
\newblock {\em Science \& Education\/}~{\em 22\/}(9), 2087--2107.

\bibitem[\protect\citeauthoryear{Achieve}{Achieve}{2011}]{Achieve2011}
Achieve (2011).
\newblock {Next Generation Science Standards}.
\newblock {\em http://www.nextgenscience.org/next-
  generation-science-standards\/}.

\bibitem[\protect\citeauthoryear{Ahlgren and Walberg}{Ahlgren and
  Walberg}{1973}]{Ahlgren1973}
Ahlgren, A. and H.~J. Walberg (1973).
\newblock {Changing Attitudes Toward Science among Adolescents}.
\newblock {\em Nature\/}~{\em 245\/}(September), 187--190.

\bibitem[\protect\citeauthoryear{Allchin}{Allchin}{2011}]{Allchin2011}
Allchin, D. (2011, May).
\newblock {Evaluating knowledge of the nature of (whole) science}.
\newblock {\em Science Education\/}~{\em 95\/}(3), 518--542.

\bibitem[\protect\citeauthoryear{Allchin}{Allchin}{2012}]{Allchin2012}
Allchin, D. (2012, July).
\newblock {Toward clarity on Whole Science and KNOWS}.
\newblock {\em Science Education\/}~{\em 96\/}(4), 693--700.

\bibitem[\protect\citeauthoryear{Allchin}{Allchin}{2013}]{Allchin2013}
Allchin, D. (2013).
\newblock {\em {Teaching the nature of science: Perspectives \& resources}}.
\newblock St Paul, MN: SHiPS Education Press.

\bibitem[\protect\citeauthoryear{Allchin}{Allchin}{2014}]{Allchin2014}
Allchin, D. (2014, January).
\newblock {From Science Studies to Scientific Literacy: A View from the
  Classroom}.
\newblock {\em Science \& Education\/}.

\bibitem[\protect\citeauthoryear{Ben\'{e}treau-Dupin}{Ben\'{e}treau-Dupin}{2013}]{BenetreauDupin2013}
Ben\'{e}treau-Dupin, Y. (2013).
\newblock {How to include the history and philosophy of science (HPS) in
  science education standards? http://www.rotman.uwo.ca/2013/02/}.
\newblock {\em Blog of the Rotman Institute of Philosophy\/}.

\bibitem[\protect\citeauthoryear{Brandt}{Brandt}{2013}]{Brandt2013}
Brandt, R. (2013).
\newblock {Broadening the Goals of Science Education
  http://www.rotman.uwo.ca/2013/02/}.
\newblock {\em Blog of the Rotman Institute of Philosophy\/}.

\bibitem[\protect\citeauthoryear{Clement}{Clement}{2009}]{Clement2009}
Clement, J. (2009).
\newblock {\em {Creative model construction in scientists and students: The
  role of analogy, imagery, and mental simulation}}.
\newblock Dordrecht: Springer.

\bibitem[\protect\citeauthoryear{Coelho}{Coelho}{2013}]{Coelho2013}
Coelho, R.~L. (2013, July).
\newblock {Could HPS Improve Problem-Solving?}
\newblock {\em Science \& Education\/}~{\em 22\/}(5), 1043--1068.

\bibitem[\protect\citeauthoryear{Conant and Nash}{Conant and
  Nash}{1957}]{Conant1957}
Conant, J.~B. and L.~Nash (Eds.) (1957).
\newblock {\em {Harvard Case Histories in Experimental Science}}.
\newblock Cambridge, MA: Harvard University Press.

\bibitem[\protect\citeauthoryear{Desimone and Porter}{Desimone and
  Porter}{2002}]{Desimone2002}
Desimone, L. and A.~Porter (2002).
\newblock {Effects of professional development on teachers' instruction:
  Results from a three-year longitudinal study}.
\newblock {\em Educational Evaluation and Policy Analysis\/}~{\em 24\/}(2),
  81--112.

\bibitem[\protect\citeauthoryear{Duschl}{Duschl}{2008}]{Duschl2008}
Duschl, R.~A. (2008, February).
\newblock {Science Education in Three-Part Harmony: Balancing Conceptual,
  Epistemic, and Social Learning Goals}.
\newblock In {\em Review of Research in Education}, Volume~32, pp.\  268--291.

\bibitem[\protect\citeauthoryear{Duschl and Grandy}{Duschl and
  Grandy}{2013}]{Duschl2013}
Duschl, R.~A. and R.~Grandy (2013, October).
\newblock {Two views about explicitly teaching nature of science}.
\newblock {\em Science \& Education\/}~{\em 22\/}(9), 2109--2139.

\bibitem[\protect\citeauthoryear{{Educational Development Center}}{{Educational
  Development Center}}{2009}]{EDC2009}
{Educational Development Center} (2009).
\newblock {Exploring Bioethics
  http://science.education.nih.gov/supplements/nih9/bioethics/default.htm}.
\newblock {\em NIH Curriculum Supplement Series Grades 9-12\/}.

\bibitem[\protect\citeauthoryear{Fox}{Fox}{2013}]{Fox2013}
Fox, C. (2013).
\newblock {What must be done to educate, equip, and support teachers to
  incorporate HPS into their curricula? http://www.rotman.uwo.ca/2013/02/}.
\newblock {\em Blog of the Rotman Institute of Philosophy\/}.

\bibitem[\protect\citeauthoryear{Garbayo}{Garbayo}{2014}]{Garbayo2014}
Garbayo, L. (2014).
\newblock {Epistemic Considerations on Expert Disagreement, Normative
  Justification, and Inconsistency Regarding Multi-criteria Decision Making}.
\newblock In M.~Ceberio and V.~Kreinovich (Eds.), {\em Constraint Programming
  and Decision Making}, Studies in Computational Intelligence 539, pp.\
  35--45. Springer International Publishing Switzerland.

\bibitem[\protect\citeauthoryear{Holton}{Holton}{1967}]{Holton1967}
Holton, G. (1967).
\newblock {Harvard Project Physics}.
\newblock {\em Physics Today\/}~{\em 20\/}(3), 31.

\bibitem[\protect\citeauthoryear{Holton}{Holton}{2003}]{Holton2003}
Holton, G. (2003).
\newblock {The Project Physics Course, Then and Now}.
\newblock {\em Science \& Education\/}~(November 2001), 779--786.

\bibitem[\protect\citeauthoryear{Holton}{Holton}{2014}]{Holton2014}
Holton, G. (2014, July).
\newblock {The Neglected Mandate: Teaching Science as Part of Our Culture}.
\newblock {\em Science \& Education\/}.

\bibitem[\protect\citeauthoryear{Holton and Brush}{Holton and
  Brush}{1985}]{Holton1985}
Holton, G. and S.~Brush (1985).
\newblock {\em {Introduction to concepts and theories in physical science}}.
\newblock Princeton, NJ: Princeton University Press.

\bibitem[\protect\citeauthoryear{Hong and Lin-Siegler}{Hong and
  Lin-Siegler}{2012}]{Hong2012}
Hong, H. and X.~Lin-Siegler (2012).
\newblock {How learning about scientists' struggles influences students'
  interest and learning in physics}.
\newblock {\em Journal of Educational Psychology\/}~{\em 104\/}(2), 469--484.

\bibitem[\protect\citeauthoryear{H\"{o}ttecke and Silva}{H\"{o}ttecke and
  Silva}{2011}]{Hottecke2011}
H\"{o}ttecke, D. and C.~C. Silva (2011, August).
\newblock {Why Implementing History and Philosophy in School Science Education
  is a Challenge: An Analysis of Obstacles}.
\newblock {\em Science \& Education\/}~{\em 20\/}(3-4), 293--316.

\bibitem[\protect\citeauthoryear{Jacquart}{Jacquart}{2013}]{Jacquart2013}
Jacquart, M.~L. (2013).
\newblock {Improving Scientific Literacy Through Improved Critical Thinking
  Skills http://www.rotman.uwo.ca/2013/02/}.
\newblock {\em Blog of the Rotman Institute of Philosophy\/}.

\bibitem[\protect\citeauthoryear{Klopfer}{Klopfer}{1992}]{Klopfer1992}
Klopfer, L. (1992).
\newblock {An historical perspective on the history and nature of science in
  school science programs}.
\newblock In {\em Teaching about the history and nature of science and
  technology}, Colorado Springs, CO, pp.\  105--130. BSCS/SSEC.

\bibitem[\protect\citeauthoryear{Klopfer and Cooley}{Klopfer and
  Cooley}{1963}]{Klopfer1963}
Klopfer, L. and W.~Cooley (1963).
\newblock {The History of Science Cases for High Schools in the Development of
  Student Understanding of Science and Scientists.}
\newblock {\em Journal of Research in Science Teaching\/}~{\em 1\/}(1), 33--47.

\bibitem[\protect\citeauthoryear{Matthews}{Matthews}{1994}]{Matthews1994}
Matthews, M.~R. (1994).
\newblock {\em {Science Teaching: The Role of History and Philosophy of
  Science}}.
\newblock Philosophy of Education Research Library. New York, London:
  Routledge.

\bibitem[\protect\citeauthoryear{{National Research Council}}{{National
  Research Council}}{1996}]{NRC1996}
{National Research Council} (1996).
\newblock {\em {National Science Education Standards}}.
\newblock Washington, DC: The National Academies Press.

\bibitem[\protect\citeauthoryear{{National Research Council}}{{National
  Research Council}}{2007a}]{NRC2008}
{National Research Council} (2007a).
\newblock {\em {Ready, set, science!: Putting research to work in K-8 science
  classrooms}}.
\newblock Washington, DC: The National Academies Press.

\bibitem[\protect\citeauthoryear{{National Research Council}}{{National
  Research Council}}{2007b}]{NRC2007}
{National Research Council} (2007b).
\newblock {\em {Taking Science to School: Learning and Teaching Science in
  Grades K-8}}.
\newblock Washington, DC: The National Academies Press.

\bibitem[\protect\citeauthoryear{{National Research Council}}{{National
  Research Council}}{2011}]{NRC2011}
{National Research Council} (2011).
\newblock {\em {A Framework for K-12 Science Education: Practices, Crosscutting
  Concepts, and Core Ideas}}.
\newblock The National Academy Press.

\bibitem[\protect\citeauthoryear{Oreskes and Conway}{Oreskes and
  Conway}{2010}]{Oreskes2010}
Oreskes, N. and E.~Conway (2010).
\newblock {\em {Merchants of doubt}}.
\newblock Bloomsbury Press.

\bibitem[\protect\citeauthoryear{Piliouras, Siakas, and Seroglou}{Piliouras
  et~al.}{2011}]{Piliouras2011}
Piliouras, P., S.~Siakas, and F.~Seroglou (2011, December).
\newblock {Pupils Produce their Own Narratives Inspired by the History of
  Science: Animation Movies Concerning the Geocentric–Heliocentric Debate}.
\newblock {\em Science \& Education\/}~{\em 20\/}(7-8), 761--795.

\bibitem[\protect\citeauthoryear{Rudge, Cassidy, Fulford, and Howe}{Rudge
  et~al.}{2014}]{Rudge2014}
Rudge, D.~W., D.~P. Cassidy, J.~M. Fulford, and E.~M. Howe (2014, January).
\newblock {Changes Observed in Views of Nature of Science During a Historically
  Based Unit}.
\newblock {\em Science \& Education\/}.

\bibitem[\protect\citeauthoryear{Rutherford, Holton, and Watson}{Rutherford
  et~al.}{1981}]{Rutherford1981}
Rutherford, F., G.~Holton, and F.~Watson (1981).
\newblock {\em {Project Physics: Resource book}}.
\newblock Holt, Rinehart and Winston Publishers.

\bibitem[\protect\citeauthoryear{Shulman}{Shulman}{1987}]{Shulman1987}
Shulman, L.~S. (1987).
\newblock {Knowledge and teaching: Foundations of the new reform}.
\newblock {\em Harvard educational review\/}~{\em 57\/}(1), 1--22.

\bibitem[\protect\citeauthoryear{Stott and Morrish}{Stott and
  Morrish}{2009}]{Stott2009}
Stott, R. and H.~Morrish (2009).
\newblock {Newton in Three Dimensions}.
\newblock {\em Science\/}~{\em 326\/}(November), 937.

\bibitem[\protect\citeauthoryear{Teixeira, Greca, and Freire}{Teixeira
  et~al.}{2009}]{Teixeira2009a}
Teixeira, E.~S., I.~M. Greca, and O.~Freire (2009, November).
\newblock {The History and Philosophy of Science in Physics Teaching: A
  Research Synthesis of Didactic Interventions}.
\newblock {\em Science \& Education\/}~{\em 21\/}(6), 771--796.

\bibitem[\protect\citeauthoryear{Welch}{Welch}{1973}]{Welch1973}
Welch, W. (1973).
\newblock {Review of the research and evaluation program of Harvard Project
  Physics}.
\newblock {\em Journal of Research in Science Teaching\/}~{\em 10\/}(4),
  365--378.

\bibitem[\protect\citeauthoryear{Welch and Walberg}{Welch and
  Walberg}{1972}]{Welch1972}
Welch, W. and H.~Walberg (1972).
\newblock {A national experiment in curriculum evaluation}.
\newblock {\em American Educational Research Journal\/}~{\em 9\/}(3), 373--383.

\end{thebibliography}

\end{document}